THE EUROPEAN
PHYSICAL JOURNAL A

# Hypernuclei production with a modified coalescence model in BUU transport calculations

Gábor Balassa[1,2,a] , György Wolf[2]

[1] Department of Physics, Korea University, Seoul 02841, Korea
[2] Institute for Particle and Nuclear Physics, Wigner Research Centre for Physics, 1525, Budapest, Hungary

**Abstract** In this paper, the usual momentum- and coordinate-space distance criteria for creating nuclear clusters in transport simulations are readdressed by using a dynamical, covariant description in an off-shell Boltzmann-Uehling-Uhlenbeck transport approach. The free parameter of this clustering scheme is the cluster formation time, which is fitted through the FOPI data of low energy charged cluster multiplicities in Au+Au collisions at 150 A MeV, and 400 A MeV incident energies. The coalescence model is used to estimate the yields of the $^3H_\Lambda$, $^5H_{\Lambda\Lambda}$, $^6He_{\Lambda\Lambda}$ single and double strange hypernuclei in central Au+Au collisions between 2, and 20 A GeV incident energies, giving comparable results to estimations from other methods.

## 1 Introduction

Hypernuclei studies are one of today's forefront research topics as they provide a very sensitive probe to study nuclear interactions, and the structure of nuclei, and could also give insight into the structure of neutron stars [1–3]. To study these kinds of nuclear systems the usual scattering studies are inadequate, due to the short lifetime of the hyperons, however spectroscopic studies of the single-, and the double strange hypernuclei could give us insight into the hyperon+nucleon, and hyperon+hyperon interactions, as well as the binding energies of the formed states [4–6]. In these studies, we consider light hypernuclei ($A = 3 - 6$) with the lightest hyperon, namely the $\Lambda$ baryon, but other hypernuclei could also be studied with the transport approach. Heavy-ion collisions are unique methods to study these kinds of systems, as the created hyperons could attach to the nucleon clusters, thus giving us the possibility to create different hypernuclei states. Measurements aiming to study these systems are carried out by e.g. STAR, and ALICE [7,8], where single hypernuclei studies are more feasible, while the double strange hypernuclei studies are mainly planned to carry out at J-PARC [9] through $K$ induced reactions, and at the CBM, and PANDA experiments at FAIR [10,11], where $p + \overline{p} \rightarrow \Xi^- \Xi^+$ reaction and its subsequent decays are used to create double hypernuclei with $\Lambda$ baryons. To estimate the expected yields multiple methods could be used such as statistical/thermal [12,13], or transport models with some well-suited coalescence criteria [14,15]. In practice two main methods are used to create nuclear fragments, each starting from different assumptions. In hybrid models, it is assumed that after the colliding stage, where the particles collide frequently, a part of the system comes close to a local or global equilibrium. After this a sudden cluster formation happens. The clusters created here then stream towards the detectors without any further interactions. This method is able to describe the multiplicities of smaller clusters in central collisions, however it is hard to give a physical explanation on how the cluster formation takes place [16,17]. The main question is that how the different clusters with low binding energies e.g. deuteron, could survive in the heath bath of $T > 100$ MeV. On the other hand many dynamical clusterization methods have been proposed, where it is assumed that the clusters are formed continuously during the time evolution, instead of a sudden clusterization, which was previously assumed in the static picture. In this case it is assumed that there is a smooth fading away of interactions, and no sudden freeze-out, which behaviour has been observed in transport calculations [18]. In theory, to be able to fully describe cluster formation it would be necessary to make full dynamical calculations, including the production and dissociation cross sections via multibody reactions, therefore treating at least the light nuclei as dynamical degrees of freedom [19]. Such calculations have been done before for deuteron formation in [20] in heavy-ion collisions using the ART hadronic transport code. More recently, the deuteron and light cluster production in heavy-ion reactions have been studied by including nucleon-, and pion catalysis reactions as well [21–23], where the multiparticle interactions are included using a stochastic collisional criteria. The dynamical approach of light cluster formation used in these studies are able to describe the measured yields and transverse momentum spectra in the examined central Au+Au and Pb+Pb collisions.

---

[a] e-mail: balassa.gabor@wigner.hu (corresponding author)

In the present work, we used our off-shell Boltzmann–Uehling–Uhlenbeck type transport code [24,25] to describe the evolution of the high-density system, and then used coalescence criteria inherited from the collision model used in our approach to describing the hyper fragments at the end of the evolution. In Sect. 2 the BUU transport model is described, with emphasis on the collisional criterion, which is later used at the coalescence stage. After summarizing the main ingredients of the transport model, the coalescence stage is described in Sect. 3, which is then later used to describe hypernuclei production in Sect. 4. At the end of Sect. 5, we summarize the method and the results.

## 2 BUU transport model and the collision criteria

Boltzmann-Uehling-Uhlenbeck type transport is a widely used approach to simulate the dynamical evolution of the dense system which is formed in heavy-ion collisions [26]. It is also suitable to describe smaller systems e.g. antiproton or pion-induced reactions, which calculations are desirable for some of the upcoming collision experiments e.g. at FAIR. The model, which will be described in this section has been developed for over a decade and is used to describe relativistic heavy-ion collisions, e.g. particle productions with success in the past [27–30]. Most recently it was extended to off-shell particle evolution and is used to examine charmonium mass shifts in antiproton-induced reactions at a few GeV center-of-mass energies, which calculations gave us a method to be able to determine the value of the gluon condensate at finite densities [31–33].

The BUU transport is developed to describe scattering up to about a few GeV bombarding energies. In the model the degrees of freedom are the hadrons, where from the baryonic side the protons, neutrons, $N(1440)$, $N(1520)$, $N(1535)$, $N(1650)$, $N(1675)$, $N(1680)$, $N(1700)$, $N(1710)$, $N(1720)$, $N(2000)$, $N(2080)$, $N(2190)$, $N(2220)$, $N(2250)$ nucleon resonances, the $\Delta(1232)$ baryon with the $\Delta(1600)$, $\Delta(1620)$, $\Delta(1700)$, $\Delta(1900)$, $\Delta(1905)$, $\Delta(1910)$, $\Delta(1920)$, $\Delta(1930)$. $\Delta(1950)$ resonances, and the $\Lambda$, and $\Sigma$ strange baryons, while on the mesonic side the $\pi$, $\rho$, $\eta$, $\sigma$, $\omega$, $K$, $J/\Psi$, $\Psi(3686)$, $\Psi(3770)$, and $D$ mesons are included as well. In the model, we do not include strings, and partonic degrees of freedom, which impose a constraint on the energy range of the model. We apply a momentum-dependent mean-field potential for the nucleons [34] according to Eq. 1:

$$U^{nr}(\mathbf{r}, \mathbf{p}) = A \frac{\rho}{\rho_0} + B \left( \frac{\rho}{\rho_0} \right)^{\tau}$$
$$+ 2 \frac{C}{\rho_0} \int d^3 p' \frac{f(\mathbf{r}, \mathbf{p}')}{1 + \left( \frac{\mathbf{p} - \mathbf{p}'}{\Lambda} \right)^2}, \quad (1)$$

where $f(\mathbf{r}, \mathbf{p})$ is the phase-space density function, $\rho_0 = 0.168$ fm$^{-3}$ is the normal nuclear density, $A = -120.3$ MeV, $B = -150.8$ MeV, $C = -64.65$ MeV, $\tau = 1.231$, and $\Lambda = 2.168$ fm$^{-1}$ free parameters are determined by fitting to the ground state properties of the nuclear medium, and to the experimentally observed momentum dependence of the one-body potential. The $\rho$ densities are calculated on fixed grid points by using Gaussian smearing and then interpolated to get the densities at specific $(x, y, z)$ points. The Lorentz invariant scalar potential is calculated from Eq. 1 by the one-particle energy as:

$$U(\mathbf{r}, \mathbf{p}) = \sqrt{\left( \sqrt{p^2 + m^2} + U^{nr}(\mathbf{r}, \mathbf{p}) \right)^2 - p^2} - m, \quad (2)$$

where $m$ is the mass of the particle, and $p$ is its momentum. In practice $U^{nr}$ is calculated in the local-rest frame, defined by the vanishing baryonic current, and it only acts on the baryons but not the mesons. For strange baryons we use different mean-field potential, see below. Apart from the nuclear mean field, we also apply a Coulomb force generated by the charged particles, which now act on mesons and baryons as well. During the evolution, Fermi-exclusion is also taken care of by calculating the $f(\mathbf{r}, \mathbf{p}, t)$ phase-space density (or simply distribution) functions on a pre-specified grid using Gaussian smearing, and then interpolating to the specific coordinate-, and momenta-points.

To initialize the system, the colliding nuclei are generated far from each other, where the potentials do not have a significant overlap. In coordinate space, the well-known Woods-Saxon distribution is used to generate the particles, while in momentum space, we generate the particles according to the local Thomas-Fermi approximation, by randomly generating momentum for each particle between 0 and $p_F(\rho)$, where $p_F$ is the density-dependent Fermi-momentum. Using this approximation the momentum distribution of the generated nuclei will have the high-energy tail, commonly found in Hartree–Fock calculations, which is important to understand subthreshold reactions in heavy-ion collisions.

Apart from the nuclear mean field and the Coulomb force, the particles could also interact with each other through 2- and 3-body collisions, which is described by a collisional integral in the transport equations. There are several collisional criteria used in different codes, however, it is not a straightforward task how to handle even the 2-body collisions in a Lorentz-invariant way [35]. In practice, it is always desirable to check the collisional criteria by comparing the results in different frames e.g. in the laboratory, and in the center-of-mass frames. In our code, we applied a collisional criterion, first described in [36], which uses a covariant formalism to check if two particles are close enough to interact or not. In this formalism the impact parameter is described as follows:

$$b = \sqrt{R_{12}^2 - \frac{h_{12}^2}{v_{12}^2}}, \qquad (3)$$

where

$$R_{12}^2 = -(x_1 - x_2)^2 - \left(\frac{p_1(x_1 - x_2)}{m_1}\right)^2 \qquad (4)$$

$$h_{12} = \frac{p_1(x_1 - x_2)}{m_1} - \frac{p_2(x_1 - x_2)m_1}{p_1 p_2} \qquad (5)$$

$$v_{12} = 1 - \left(\frac{m_1 m_2}{p_1 p_2}\right)^2, \qquad (6)$$

where $x$ is the space-time coordinate, $p$ is the four-momenta, and $m$ is the mass of the particles. Other than the center-of-mass system of the colliding particles, there are two corresponding times, that belong to the collision. In the eigensystem of the two particles, these can be expressed as:

$$\tau_1 = -\frac{p_1(x_1 - x_2)}{m_1} + \frac{h_{12}}{v_{12}^2} \qquad (7)$$

$$\tau_2 = \frac{p_2(x_1 - x_2)}{m_2} + \frac{h_{21}}{v_{12}^2} \qquad (8)$$

In our approach, we define the collision time in the system used in the calculations as:

$$dt = \frac{1}{2}\left(\frac{e_1}{m_1}\tau_1 + \frac{e_2}{m_2}\tau_2\right), \qquad (9)$$

where $e$ is the energy of the particle. For a specific reaction to being able to take place $dt$ has to be smaller than the time step. The main problem is causality, which is impossible to impose on space-like events. Causality violations will cause non-invariant time ordering, which has to be at least minimized in practical calculations. For time-like events, it is possible to conserve time-ordering by imposing a causality condition for the proper time intervals between two collisions and perform only collisions, which are next in line for each particle [36]. In this way, causality violation can be minimized and this is usually enough for practical calculations. Our method shows only a few percentage differences between the collision numbers observed in the laboratory and center-of-mass frames in 400 A MeV and 2.1 A GeV Ca+Ca collisions.

Using the described method, two particles are interacting with each other if $b \leq \sqrt{\sigma/\pi}$, where $\sigma$ is the cross section of the specific reaction, and $b$ is the impact parameter. The time constraint in Eq. 9 also has to be satisfied for a reaction to be able to take place. In our code, we consider elastic baryon+baryon, inelastic baryon+baryon, meson+meson, and inelastic meson+baryon collisions for all of the baryons, and mesons, which are included in the code. Apart from collisions, the unstable particles could also decay during their propagation, which is described by their decay width through the probability $P = 1 - e^{-\Gamma/\hbar\gamma\,\Delta t}$, where $\Delta t$ is the timestep of the simulation, $\gamma$ is the Lorentz factor, while $\Gamma$ is the decay width of the resonance, which for e.g. the $\Delta_{1232}$ resonance can be energy dependent as well. A detailed description of the base model can be found in [27], while some recent extensions with charmonium production cross sections and off-shell propagation can be found in [31].

To make nuclear fragments, we will use the above-described collisional criteria, with a small extension, by considering the formation time of the nuclear fragments. In this way, the coalescence model is strongly connected to the actual method of how the particles can interact in our approach.

## 3 Coalescence model

In cascade models, a coalescence stage can be applied after freeze-out, where it is usually used to make deuterons, tritiums, light-, and heavier fragments, from the neutrons, and protons available at the end of the reaction [37,38]. To create the fragments static, and dynamical approaches could be used, where the former method assumes a sudden cluster formation after the system reaches a local or global equilibrium, while the latter assumes a smooth fading away of interactions, while the clusters could be formed continuously during the time evolution. One of the most basic coalescence model is when a composite particle is formed according to the position and relative momenta of the participating nucleons. Checking the relative momenta and coordinate distances between the nucleons in their center-of-mass frame, spanning trees can be built, which contain all the possible k-body subclusters, which are built up by all of the participating nucleons. This method proved successful to describe the multiplicities of e.g. deuteron, and triton formation by fitting the momentum ($p_c$), and coordinate ($r_c$) radius [39], however, there could be some inconsistencies with the rapidity distributions, and in the low energy yields as well. In particular, the low energy yields of the $^4$He and $^3$He particles show an interesting behavior, as the $^4$He multiplicity is larger, than the $^3$He yield below a few hundred of A MeV incident energies, which could not be described by simple coalescence models [40]. The ability to describe such processes is essential to better understand the physics behind fragment formation in nuclear physics. In practice to describe the different types of fragments, the free parameters could be different as well. Another model, which is used to create nuclear fragments is called the simulated annealing clusterization algorithm (SACA method) [41,42], which finds the most stable clusters by examining the binding energies of the possible cluster patterns, or the FRIGA method [43], which includes symmetry, and pairing energy in the cluster formation as well. In other dynamical models, the created fragments are considered as excited states after their formation and have to be de-excited to create the final stable clusters [44,45]. All of these methods could be used to describe hypernuclei formation as well, by including hyperons in the coalescence stage. It is not straightforward, however, what should be the free parameters of the aforementioned models. In principle, these should take into consideration the interaction strengths between the various particles, which are not well-known for e.g. the hyperon-nucleon, or hyperon-hyperon interactions.

In our approach, we aim to avoid the problem of 'not unique' coalescence parameters (at least partially), by extending the collision criterion, described in Sect. 2, to create light fragments after freeze-out. To validate the model, in this section only the conventional fragments, which are built up by protons and neutrons are considered, while as our ultimate goal is to describe hypernuclei, in the next section we extend the model by adding hyperons into the picture as well.

Starting from the collisional criteria, firstly, we assume that two particles can interact with each other if their respective impact parameter in Eq. 3 is smaller than $\sqrt{\sigma(\sqrt{s})/\pi}$, where $\sigma(\sqrt{s})$ is the energy-dependent cross section of the elastic two-body scattering. In practical calculations the elastic cross section will coincide with the total cross section as the energies of the particles in the final stage will be lower than the first inelastic threshold. This in itself, however, is not enough, as we also need to satisfy $dt \leq \Delta_C$, where $dt$ is defined through Eq. 9, and $\Delta_C$ is a free parameter, called the coalescence time. Furthermore, we assume that $\Delta_C$ is unique, and has to be fitted from experiments. In simple instantaneous collisions, $dt$ had to be smaller, than the timestep of the simulation ($\Delta t$), however, here, we let the particles interact for a (possibly) longer time to create the fragments.

Apart from the interaction criteria, we also check how stable are the interacting particles, by letting them interact at least for $10\Delta t$ timesteps. If the created clusters will stay the same for the whole time interval, they will have a probability of 1 to stay together. In contrast, if they break apart in this time interval, they will get a penalty factor, which will suppress their probability to be created. In our model, we count down the times that a specific cluster is formed in the last 10 timesteps, and multiply the final multiplicities by a factor $n_i/10$, where $n_i$ is the number of appearances of a specific cluster in the last 10 timesteps. In practice, if we let the system evolve for a long time corresponding to the few GeV energies ($t \approx 50$–100 fm/c), the clusters will be stable almost every time, and the penalty factor only adds a little fluctuation to the final results. It is a necessary step to check the stability of the created clusters because an instantaneous criterion could lead to clusters that could be broken up a few timesteps ahead. In [46] a similar description is used, where a persistence coefficient is introduced to measure the tendency of the interacting particles to stay together in a specific nuclear cluster. In [47] the cut-off time dependence of the yields, and transfer momentum spectra of the p, n, d,$^3$He, and $^4$He particles are examined using the parton-hadron-quantum-molecular-dynamics (PHQMD) approach, where the results show a slight, but observable decreasing tendency in the final yields with higher cut-off times $\sim 50$–70 fm/c in central Au+Au collisions at midrapidity. Furthermore in [48] the production of deuterons are studied with a coalescence model (where the deuterons are created at the kinetic freeze-out hypersurface), and with the minimum spanning tree (MST) method by taking into account the cut-off times used in the ultrarelativistic quantum molecular dynamics (UrQMD) and PHQMD transport approaches. In that study, the transverse distance of the free protons, neutrons, and the bound deuterons are calculated at different cut-off times from 30 to 70 fm/c in central Pb+Pb collisions, showing that the deuterons tend to stay together at a smaller radial distance than the free nucleons. This result is in contrast with the ones obtained by the statistical model assumptions, where a homogenous distribution of nucleons and deuterons are assumed. An interesting implication of these results could be that in this case the deuterons do not pass through the hot fireball and stay somewhat seperated from it, which would explain why they are not destroyed in the hot medium. In our method the cut-off time dependence is avaraged out by the previously described penalty factor.

The interaction criterion is similar to the simple coordinate-, and momentum-space closeness criterion. The main difference is that in this case the coordinate-space closeness parameter directly depends on the energy and the participants through the energy-dependent cross sections, which are used to calculate the impact parameters, and is not a free parameter. In the original coordinate-, and momentum-space coalescence picture the $\Delta p$ momentum parameter makes sure that the participating particles approximately move towards the same direction, with around the same speed, so they could stay together for a longer time. In our picture this is satisfied through the $\Delta_C$ coalescence time, which is a parameter in the Kodama collisional criterion, which in this case has the physical meaning, that the particles could stay close together for some time, making it possible them to make a stable cluster. Therefore our description, in principle, is very similar to the basic coalescence model, however, operates with different free parameters. The main difference is the dependence of the coalescence stage on the participants and their cross sections, and it will be interesting to compare the differences it could make in the resulting clusters.

As the interaction criterion consists of the elastic cross sections of the different interactions, it is necessary to include those from measurements or model calculations. In the low-energy region for protons and neutrons, these cross sections are relatively well measured, which is necessary because the thermalized particles will have relatively low momentum at

 

the end of the evolution of the system. For the reactions $pp$, $nn$, and $pn$ the well-known Cugnon parametrization is used [49], which is also used for the instantaneous collisional criteria in our code. Using the above-described model, with the different cross sections, the only free parameter is now the $\Delta_C$ coalescence time, which is fitted through the low energy charged cluster multiplicities, measured at the GSI, with the FOPI detector [40,50]. In these calculations, we used the ERAT criterion, defined in Eq. 10 for event selection:

$$ERAT = \frac{\sum_i p_{i,T}^2/(m_i + E_i)}{\sum_i p_{i,L}^2/(m_i + E_i)}, \tag{10}$$

where $p_T$ is the transverse momenta, $p_L$ is the longitudinal momenta, $E_i$ is the energy, while $m_i$ is the rest mass of the fragments. To be able to compare our calculations to the measurements, the fragment masses are defined as $m_i = 2Z_i m_N$, where $Z$ is the nuclear charge, $m_N$ is the nucleon mass, and the ERAT value is calculated at the forward hemisphere. In our calculations, we used the centrality cut ERAT > 0.6, which corresponds to the FOPI measurements and can be done easily in the transport simulations. The main disadvantage of using such an event selection criterion is that it needs many simulations if we want to select only the most central events. In these specific cases, approximately a few percent are selected from the total of 1 million simulated events. Recent theoretical works that are describing the FOPI data can be found in [51,52], where it is assumed that the primary diluted clusters in the low-density baryonic matter will undergo statistical decay leading to the final nuclear fragments.

To fit the $\Delta_C$ free parameter, we calculated the average relative error defined in Eq. 11 for the charged fragment multiplicities in Au+Au collisions up to $Z = 8$ at 150 A MeV, and 400 A MeV bombarding energies. The results shown in Fig. 1 indicate that a $\Delta_C \approx 28$ fm/c value for the coalescence time is the best value to describe the data, with reasonable accuracy.

$$\mathcal{E}_{REL} = \frac{1}{16} \sum_{J=1}^{2} \sum_{i=1}^{8} \frac{|n_{Z=i}^{meas,J} - n_{Z=i}^{model,J}|}{|n_{Z=i}^{meas,J}|}, \tag{11}$$

where the index $J$ corresponds to the bombarding energy, which in this case could take two values 150 A MeV, and 400 A MeV. Furthermore, $n_{Z=i}^{meas/model,J}$ corresponds to the measured/calculated multiplicity of the $Z = i$ charged fragments at the corresponding $J$'th energy.

The results for the charged fragment multiplicities at 150 A MeV, and 400 A MeV bombarding energies can be seen in Fig. 2.

To compare our method to the coordinate-, and momentum space coalescence, we have made the calculations with that method as well by fitting the $\Delta r$ parameter to 4 fm, and using the corresponding $\Delta p = \hbar/\Delta r$ momentum space distance to create the clusters. The results for 400 A MeV bombarding energy can be seen in Fig. 3, where a very good match is achieved between the two methods.

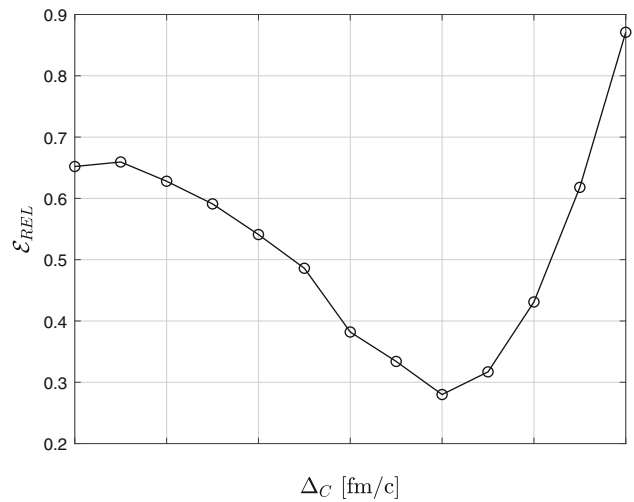

**Fig. 1** The avaraged relative error for the charged multiplicities up to $Z = 8$ in Au+Au collision. The minumum value corresponds to $\Delta_C = 28$ fm/c

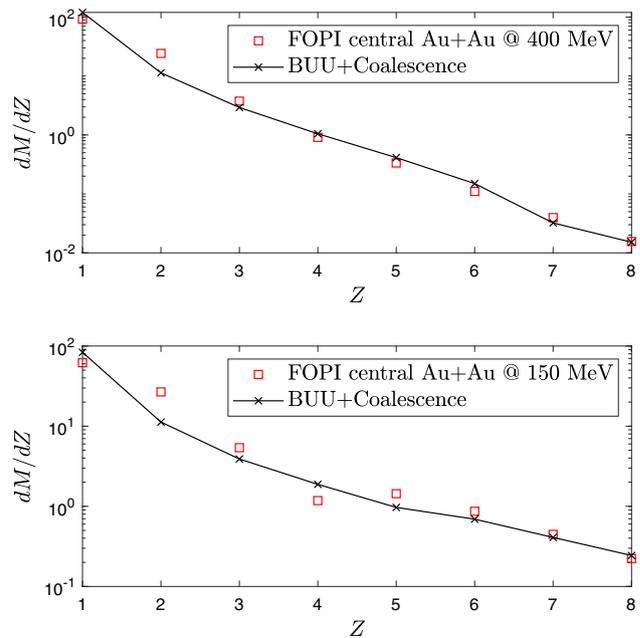

**Fig. 2** Validation of the coalescence model through the low energy FOPI data [50] for the charged fragment multiplicities, using $\Delta_C = 28$ fm/c

Apart from the charged particle multiplicities, it is also important to check other observables e.g. the kinetic energy distribution of the created fragments. To this purpose, we have calculated the average kinetic energy dependence on the mass number $A$ of the fragments and compared them with the FOPI measurements at 150 A MeV in the polar angle range 25°–45°. The results can be seen in Fig. 4.

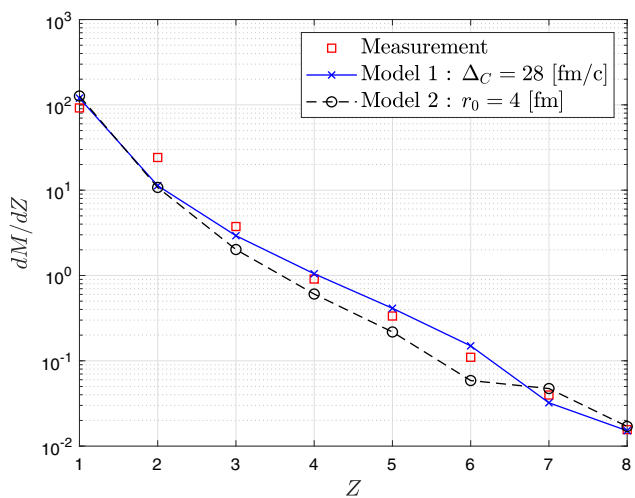

**Fig. 3** Comparison of the simple coordinate-, and momentum space distance coalescence, and the method described in this paper, using the FOPI charged multiplicity data at 400 A MeV bombarding energy Au+Au collisions

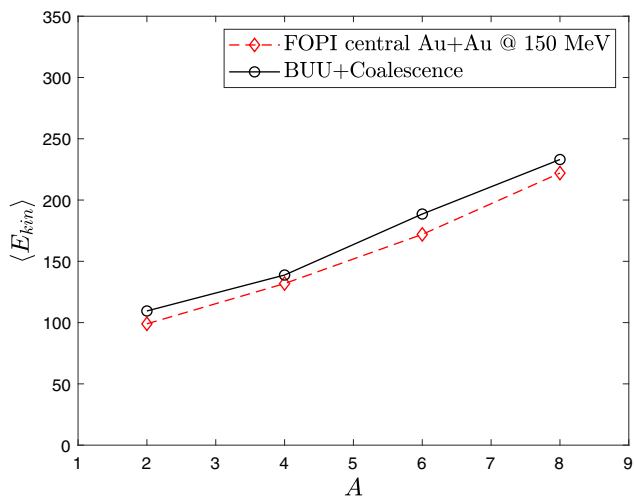

**Fig. 4** Avarage kinetic energies of the fragments, calculated by our model, and compared to the FOPI data at 150 A MeV in the polar angle range 25°–45°

One very important experimental result, which helps better understand the underlying physical processes in fragment formation is the behavior of the relative yields of the $^3$He, and $^4$He nuclei. Simple coalescence models predict slightly larger $^3$He yields at a few hundreds of A MeV energies, however, measurements tell us that e.g. in Au+Au collisions approximately under 500 A MeV incident energy, the $^4$He yields are larger, than the $^3$He yields and there is a crossing point between 400 and 600 A MeV. As was mentioned before, this energy dependence cannot be described by a simple coalescence picture, and our coalescence method also fails to reproduce this behavior as it gives slightly larger $^3$He yields at incident energies of a few hundred A MeV in central

Au+Au collisions. An important sidenote is that this behavior can be explained by introducing statistical a multifragmentation model after the coalescence step, which starts from the assumption that the produced fragments are excited, and have to undergo a secondary de-excitation. This way the different isotopes are modified (especially the larger ones) through the de-excitation process, and it has been shown in [53], that e.g. the low energy $^3$He, and $^4$He yields can be explained. In future studies, we would also like to include statistical multifragmentation into our model after the coalescence step, however, in this paper, only the basic coalescence model is used to describe fragment formation. In the next section the model will be used to describe hypernuclei production at a few A GeV bombarding energies.

## 4 Hypernuclei production

In this section, the hypernuclei production will be described by using the coalescence model shown in the previous section. Until now only nucleons took part in the coalescence stage, however, if one wants to create hypernuclei it is necessary to include the different hyperons as well. The BUU transport code includes near and subthreshold strangeness production for the $\Lambda$, $\Sigma$, and $\Xi$ baryons, as well as for $K$ meson production [54–57]. In our approach we employ a weakly repulsive momentum-dependent mean field for the $K$ mesons, and momentum-dependent in-medium single-particle potentials to the $\Lambda$, $\Sigma$, and $\Xi$ baryons. The parameters for the strange baryon potentials are extracted from Lattice QCD calculations [58]. For the $\Lambda$ baryon, the potential is assumed to be proportional to the nucleon in-medium potential $U_\Lambda = 0.47U_N$, which form is a good approximation of the momentum dependence found at normal nuclear density [54,58]. To create the $\Lambda$, and $\Sigma$ hyperons, we have used an effective Lagrangian model to calculate the elementary cross sections in $N + N$, $N + \Delta$, and $\Delta + \Delta$ collisions [59]. During the evolution of the system, the $\Lambda$, and $\Sigma$ hyperons could be absorbed on nucleons through the reaction $\Lambda(\Sigma) + N \rightarrow NN\overline{K}$, which is also implemented into the transport approach, and is used to study antikaon production in [60] before. Apart from the creation and absorption of hyperons, the elastic interactions with nucleons is also considered. A more detailed description of hyperon studies with the BUU transport approach can be found in [54].

In these calculation, we will only consider $\Lambda$ hyperons, which could join a nucleon cluster to create hypernuclei with one or two $\Lambda$ inside. Following the coalescence criteria, we need the elastic cross sections for $\Lambda N$, and $\Lambda\Lambda$ interactions, which are not well known. To estimate these cross sections, we could use e.g. effective model calculations [61,62], however, for the first estimation, it is not necessary as there are some measurement data for the $\Lambda N$ interaction, which can be used for creating hypernuclei with one $\Lambda$ inside [63–67]. For double strange hypernuclei, where we have two $\Lambda$-s the $\Lambda\Lambda$ interaction could also play a role. In this case, for the first estimation, we will assume that the $\Lambda\Lambda$ interaction is approximately the same as it is in the $\Lambda N$ case, however, we also make calculations with a suppressed $\Lambda\Lambda$ interaction strength to see the difference it could make in the final multiplicities. In Fig. 5 the multiplicities per event can be seen for the $^3\text{H}_\Lambda$, $^5\text{H}_{\Lambda\Lambda}$, $^6\text{He}_{\Lambda\Lambda}$ single and double hypernuclei in Au+Au collisions, with 0–10% centrality at different bombarding energies, using the assumption that $\sigma_{N\Lambda}(\sqrt{s}) = \sigma_{\Lambda\Lambda}(\sqrt{s}-\Delta M)$, where $\Delta M = (2m_\Lambda) - (m_N + m_\Lambda)$ is the corresponding mass shift due to the different thresholds. The results show the same order of magnitude, and energy dependence as other studies, where transport simulations with coalescence models [68], and thermal models [69,70] were used to estimate the hypernuclei yields in heavy-ion collisions.

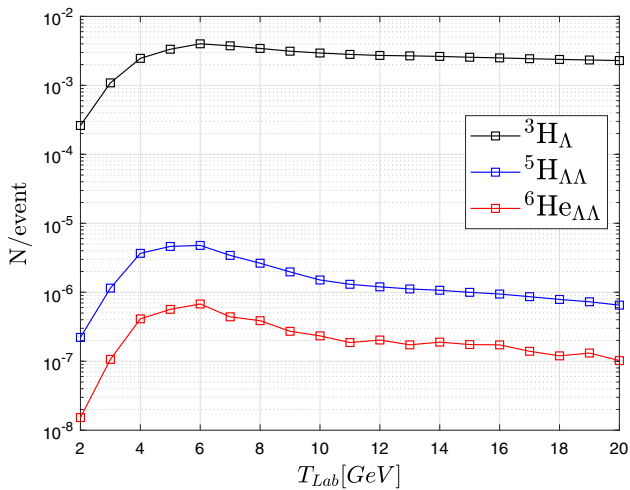

**Fig. 5** Hypernuclei production yields in Au+Au collisions at $0 - 10\%$ centrality, using the assumption of equal $N\Lambda$ and $\Lambda\Lambda$ interaction strengths

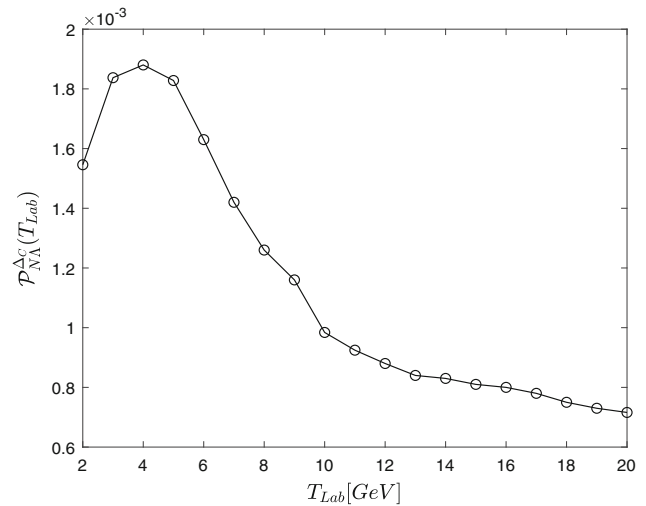

**Fig. 6** The elementary $N\Lambda$ coalescence probability $\mathcal{P}_{N\Lambda}^{\Delta_C}(T_{lab})$ in central Au+Au collisions

In all cases, the yields for the hypernuclei show a maximum of around 4–6 GeV laboratory energies, and a slow decreasing part after that. In principle, this behavior corresponds to the relative yields, and momentum distributions of the hyperons and nucleons after kinetic freeze-out, which ultimately depends on the elementary cross sections, which are used to create and scatter these particles. In our model as we use the elastic $NN$, and $N\Lambda$ cross sections to create the clusters such an energy dependence is expected. To study this behavior further, we can examine the probability, that a $\Lambda$ hyperon and a nucleon coalesce together in a specific collisional setup for example in this case 0-10% central Au+Au collisions. These probabilities should depend on the system under study e.g. on the energy of the collision, the centrality, and the colliding nuclei, so the results shown here will be unique to this specific collisional configuration. In principle, this probability can be described by the probability distribution function $\mathcal{F}(T_{lab}, b, x_1, x_2, p_1, p_2, m_1, m_2, \Delta_C)$, where $T_{lab}$ is the laboratory kinetic energy, and $b$ is the impact parameter of the heavy-ion collision, $x_i$ is the space-time coordinate, $p_i$ is the four-momentum, $m_i$ is the mass of the i'th particle, and $\Delta_C$ is the free parameter of the coalescence method. Here, we are only interested in the integrated probabilities, with fixed masses, and fixed $\Delta_C$ as:

$$\mathcal{P}_{N\Lambda}^{\Delta_C}(T_{lab}) = \int_{0-10\%} db \int dx_1 dx_2 dp_1 dp_2 \\ \times \mathcal{F}_{m_1, m_2, \Delta_C}(T_{lab}, b, x_1, x_2, p_1, p_2), \quad (12)$$

where $\mathcal{P}_{N\Lambda}^{\Delta_C}(T_{lab})$ is the integrated probability for the elementary $N\Lambda$ coalescence in Au+Au collisions at 0-10% centrality. The fixed parameters $m_1$, $m_2$, and $\Delta_C$ are shown in the subscript of the probability distribution function to make a clear distinction from the other variables. The final probability could be approximated through the BUU transport and the results can be seen in Fig. 6.

The distribution clearly shows the observed energy dependence and the maximum around 4 GeV, however, the full picture is more complicated than just observing the elementary coalescence probability of the $\Lambda$ baryons and the nucleons, as this function only tells us the probability that a nucleon and a hyperon will be close enough to be joined together. For a specific cluster, we also need that the particles only interact with each other, and no one else outside their cluster, which also gives us an extra contribution with a different energy dependence. The final functional form also depends on the $NN$ coalescence probability, which is different than the $N\Lambda$ one due to their different thresholds, cross sections, and momentum distributions. The final yields in Fig. 5 will depend on all of these contributions. It is also worth noting that the decrease of the hypernuclei yields are very slow after

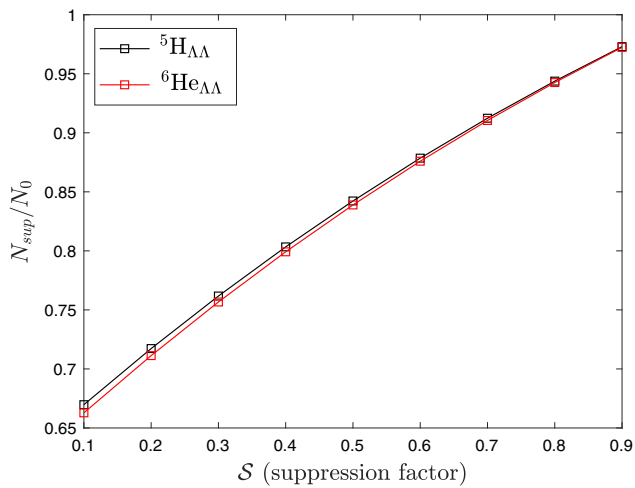

**Fig. 7** Dependence of the $^5H_{\Lambda\Lambda}$, and $^6He_{\Lambda\Lambda}$ hypernuclei yields on the $\Lambda\Lambda$ suppression parameter $\mathcal{S}$ (defined in Eq. 13) at $T_{Lab} = 6$ GeV, where $N_0$ is the hypernuclei yield with $\mathcal{S} = 1$, and $N_{sup}$ is the yield with $\mathcal{S} < 1$. Lower $\mathcal{S}$ values mean larger suppression of the $\Lambda\Lambda$ interaction compared to the $\Lambda N$ interaction, in which case the difference in the yields will be more significant

the maximum values near $4 - 6$ A GeV laboratory kinetic energies, which means FAIR and NICA energies are optimal to measure these states [70].

To further study the yield dependence on the $\Lambda\Lambda$ interaction strength, we also calculated the yields with different $\sigma_{\Lambda\Lambda}$ cross sections. For simplicity, we assumed that the energy dependence of the $\Lambda\Lambda$ cross section is equal to the $N\Lambda$ cross section, apart from an energy shift due to the different thresholds, while the magnitude is suppressed by a constant factor as:

$$\sigma_{\Lambda\Lambda}(\sqrt{s}) = \mathcal{S} \cdot \sigma_{N\Lambda}(\sqrt{s} - \Delta M), \quad (13)$$

where $\Delta M = (2m_\Lambda) - (m_N + m_\Lambda)$ is the mass shift due to the different thresholds, while $\mathcal{S}$ is a suppression factor, which in our simulations is set to go from 0.1 to 0.9 with $\Delta \mathcal{S} = 0.1$. The results for the two double strange hypernuclei at $T_{Lab} = 6$ GeV are shown in Fig. 7.

From these results, it can be seen that the final suppression in the yields is quite small, and even if the $\Lambda\Lambda$ interaction is $1/10$'th of the $N\Lambda$ interaction, the final yields will give the same order of magnitude estimates. It is also worth noting that in all of these cases, there is no significant difference in the final yields for the $^5H_{\Lambda\Lambda}$, and $^6He_{\Lambda\Lambda}$ hypernuclei. The results shown here are the consequence of the specific coalescence method we used to estimate the hypernuclei yields, therefore it could be an important test to check the validity of the model.

## 5 Conclusions

Studying nuclear fragments in heavy-ion collisions is an important task to understand the interaction between the constituents of nuclear matter. These studies are important not just in understanding the strongly interacting matter, but to understand astrophysical processes as well. There are several methods that can describe nuclear fragment formation in heavy-ion collisions, each having some success, especially in describing small fragments, however, there is no ultimate method on how to handle the coalescence of nucleons and other particles. In this paper, we proposed a coalescence model, based on the Kodama collisional criteria, which is used in our off-shell Boltzmann–Uehling–Uhlenbeck transport code. In the model, we use the elastic cross sections of the possible interactions, if the participants are interacting with each other or not, and leaving the time of interaction (coalescence time) as the only free parameter. The coalescence time is fitted through the low energy FOPI data for charged fragments and is used to describe hypernuclei production in a moderate $T_{Lab} \in [2, 20]$ GeV energy range in 0–10% central Au+Au collisions. The estimated yields for the $^3H_\Lambda$, $^5H_{\Lambda\Lambda}$, and $^6He_{\Lambda\Lambda}$ hypernuclei are in good agreement with previous estimates from different models. By varying the $\Lambda\Lambda$ interaction strength through a suppression factor, we have estimated the dependence of the final yield on the suppression factor. According to these results the final yields of the observed double hypernuclei will stay at the same order of magnitude even if the $\Lambda\Lambda$ interaction is $1/10$'th of the $N\Lambda$ interaction. In future studies, we intend to include other single and double strange hypernuclei, with more emphasis on their rapidity distribution as well.

**Acknowledgements** This work was supported by the National Research Foundation of Korea (NRF) grant funded by the Korea government(MSIT) (No. 2018R1A5A1025563). The authors were supported by the Hungarian OTKA fund K138277, and the EU project EURIZON.

**Funding Information** Open access funding provided by ELKH Wigner Research Centre for Physics.

**Data Availability Statement** This manuscript has no associated data or the data will not be deposited. [Authors' comment: All the data shown in the paper are available from the corresponding author, upon request].